\newcommand{\appsection}[1]{\let\oldthesection\thesection
  \renewcommand{\thesection}{Appendix \oldthesection}
  \section{#1}\let\thesection\oldthesection}

\documentclass[12pt,fleqn,psfig]{article}
\usepackage{amsmath, amssymb, graphicx, float,appendix}

\begin{document}


\textwidth 6.3in \textheight 8.8 in \hoffset -0.5 in \voffset -0.3
in
\renewcommand{\theequation}{\thesection.\arabic{equation}}


\thispagestyle{empty}
\renewcommand{\thefootnote}{\fnsymbol{footnote}}

{\hfill \parbox{4cm}{
        Brown-HET-1620 \\
}}

\bigskip\bigskip\bigskip\bigskip\vskip100pt

\begin{center} \noindent
On the Asymptotics of the Hopf Characteristic Function
\end{center}

\bigskip\bigskip\bigskip

\centerline{ \normalsize  Zachary Guralnik$^{1}$\footnote[1]{zach@het.brown.edu}, Cengiz Pehlevan$^{2,1}$\footnote[2]{cengizpehlevan@fas.harvard.edu},  Gerald Guralnik$^1$\footnote[1]{gerry@het.brown.edu} }

\bigskip

\bigskip

\centerline{1. Department of Physics}
\centerline{Brown University}
\centerline{Providence, RI 02912}
\bigskip

\centerline{2. Harvard University}
\centerline{Center for Brain Science}
\centerline{Cambridge MA, 02138}
\bigskip


\bigskip\bigskip\bigskip\bigskip

\renewcommand{\thefootnote}{\arabic{footnote}}

\centerline{ \small Abstract}
\bigskip

\small
We study the asymptotic behavior of the Hopf characteristic function of fractals and chaotic dynamical systems in the limit of  large argument. The small argument behavior is determined by the moments, since the characteristic function is defined as their generating function.   Less well known is that the large argument behavior is related to the fractal dimension.  While this relation has been discussed in the literature, there has been very little in the way of explicit calculation.  We attempt to fill this gap, with explicit calculations for the generalized Cantor set and the Lorenz attractor.   In the case of the generalized Cantor set, we define a parameter characterizing the asymptotics which we show corresponds exactly to the known fractal dimension.  The Hopf characteristic function of the Lorenz attractor is computed numerically,  obtaining results which are consistent with Hausdorff or correlation dimension, albeit too crude to distinguish between them.

\newpage

{\bf

Because of the impossibility of predicting the detailed behavior of chaotic dynamical systems over long times, one is generally only interested in their geometric and statistical properties.  Nevertheless, the predominant means of computing these properties is by direct numerical simulation of a trajectory which is subsequently analyzed like experimental data.  While this is a viable approach,  it is not formulated in terms of the quantities of interests and therefore yields little insight.  Moreover, direct simulation is probably not the most efficient approach, since the simulated trajectory contains much more information than can be reliably calculated.  Fortunately, the equations of motion yield equations for statistical quantities, such as the Hopf characteristic function $Z(\vec J)$, which might be soluble without the need for direct simulation of a chaotic trajectory.

Associated with a chaotic trajectory is an invariant measure $\mu$ which determines the frequency with which the system may be found in a region of phase space. The Hopf characteristic function is the Fourier Stieltjes transform of this measure,
$
Z(\vec J)\equiv\int d\mu\, \exp(i\vec J\cdot\vec x)
$
The small $J$ behavior of the Hopf characteristic function determines the equal time correlation functions of a chaotic attractor, via the Taylor--Maclaurin series expansion in $\vec J$.

In this article, we will study the large $J$ asymptotics of the characteristic function.  While small $J$ behavior is related to the correlation functions,   the large $J$ behavior is related to the geometry of the attractor, specifically its dimensionality. 
The geometric structure of chaotic attractors, and their fractal cousins, is bizarre to the uninitiated.  Among the strange but commonly observed features is non--integral dimensionality, and fine structure at all scales.  With these features comes the attendant impossibility of defining  a finite distribution function on the attractor or fractal.  
The absence of a distribution function means  that the measure $d\mu$ can not be written as $\rho(\vec x)d\vec x$ for any function $\rho(\vec x)$. In this case the Fourier transform of the characteristic function, $\int d\vec J Z(\vec J)\exp(i\vec J\cdot \vec x)$, can not converge.  The associated constraint on the  asymptotic behavior of $Z$ underlies quantitative statements which can be made relating the dimensionality to the asymptotics of the characteristic function.

The relationship between fractal measures and the asymptotics of their Fourier-Stieltjes transform has been discussed in the literature \cite{KPG}, however there has been very little in the way of explicit calculation.  The intent of this article is to fill this gap. The Hopf characteristic function can be defined for a fractal,  with asymptotic properties which are often readily computable due to the self similarity which is a hallmark of fractals. We will determine  asymptotics properties of $Z(\vec J)$ for the generalized Cantor set, showing the correspondence with the known fractal dimension.  Our arguments can be readily generalized to other fractals.

Determing the asymptotics of the Hopf function of a chaotic dynamical system is a much more difficult, and mostly unsolved, problem. We also attempt to determine the asymptotics of the characteristic function of the Lorenz attractor numerically.  The results for the Lorenz attractor are crude, being limited to large but finite $\vec J$, yet seem to be consistent with the known dimensionality.  It is our hope that this crude approach may be supplanted by one in which the asymptotics is determined from the differential equations which the Hopf characteristic function satisfies, subject to the appropriate boundary conditions.
}

\section{Introduction}

Chaotic systems are generally studied by direct numerical simulation, with statistical (e.g. moments) and geometric (e.g. dimension) information determined from a time series.
However, because of senstitivity to initial conditions, the detailed dynamics of any given time series is rarely of interest.  An alternate approach, formulated entirely in terms of the statistical quantities of interest, is to attempt to solve the
equations and boundary conditions which determine the Hopf characteristic function $Z(\vec J)$ (see \cite{Hopf,Frisch}.  Recently this approach has been applied to problems in atmospheric dynamics and stellar physics \cite{Marston1, Marston2}).

For  a  dynamical system with phase space variables $X_1,\cdots,X_n$,  the characteristic function $Z(\vec J)$ is defined  \cite{Hopf,Frisch}  as the time average of \nobreak $\exp\left(i\vec J\cdot \vec X(t)\right)$.
Assuming ergodic behavior,
\begin{align}
Z(\vec J)\equiv\left< \exp\left(i{\vec J}\cdot{\vec X(t)}\right) \right>=\int d\mu(\vec X)\exp\left(i{\vec J}\cdot\vec X\right)\,,
\end{align}
where $d\mu(\vec X)$ is the formal representation of the invariant measure.
For a fractal which is not derived from a dynamical system, the characteristic function is defined solely in terms of a choice of measure $d\mu(\vec X)$.  The small $J$ behavior of $Z$ carries information about
the moments $\left<X_i\cdots X_j\right>$, via the Taylor--Maclaurin expansion.  Remarkably, and less well known, is that the large $J$ behavior carries information about the dimensionality of the chaotic orbit or fractal. We will be concerned with the large $J$ behavior in this article.

When the invariant measure can be written in terms of a probability density function, $d\mu(\vec X)=\rho(\vec X)d^n X$,  where $\rho(\vec x)$ is everywhere finite, $Z(\vec J)$ and $\rho(\vec X)$ are Fourier transforms of each other. However in many systems of interest, the dynamics is dissipative and a probability density function which is everywhere finite does not exist, although it may exist as a distribution. If an everywhere finite $\rho$ did exist, it would satisfy $\frac{d\rho}{dt} = -\rho \vec\nabla\cdot\vec v >0$ along any trajectory, where $\vec v(\vec X)$ is the velocity $\frac{d\vec X}{dt}$, which is inconsistent with Poincare recurrence.
In the presence of dissipation, the dimensionality $d$ of the chaotic orbit  is less than dimension $n$ of the space spanned by $\vec X$, and may in fact be
non-integer.  For a dissipative system with  $d<n$,  the Fourier transform of $Z(\vec J)$ can not converge. A sufficient condition for the existence of a Fourier transform is that $Z(\vec J)$ be ${\cal L}^2$, such that the integral $\int d^n J\, Z(\vec J)^*Z(\vec J)$ converges.  Thus it would seem natural to define a dimensionality corresponding to the maximum value of  $s\le n$ such that the integral
\begin{align}\label{defcon}I_s=\int_{|J|>\epsilon} d^n J\, |J|^{s-n} |Z(\vec J)|^2
\end{align}converges.

The characteristic function can be given exactly for some systems with integer dimension $d$, such as the Orszag--Mclaughlin model \cite{MarstonMa}.  In this and any other case in which the invariant measure can be written as $\rho(\vec X)d^n X$ where $\rho$ is product of finite functions with delta functions, it is not hard to see that the maximum value of $s$ of which $I_s$ converges is indeed equal to the dimension of the chaotic trajectory; $s_{\rm max}=d$.
The asymptotics of the characteristic function is not so trivially determined when the dimension is fractional.
In fact, $Z(\vec J)$ need not fall off at large $J$, except in some averaged sense \cite{PerSjolin,Erdogan}.
It is known that the maximum value of $s$ for which $I_s$ converges is bounded above by the Haussdorf dimension $d$.
$I_s$ is proportional to the
`s-dimensional energy of the measure', defined by
\begin{align}
E_s(\mu)=\int\int\frac{ d\mu(\vec X) d\mu(\vec Y)}{|\vec X-\vec Y|^s}\, ,
\end{align}
the inverse of which is known as the s-capacity.  For all measures on a  Borel set of dimension $d$ in a complete metric space,  the s-capacity  vanishes for $s>d$.  For $s\le d$ there exists a Borel measure such that the s-capacity is finite \cite{Edgar}.  In this article we explicitly evaluate the asymptotic behavior of the Hopf characteristic function, or convergence criteria for $I_s$, for some well known chaotic dynamical systems and fractals.

The relation between fractal measures and the asymptotics of their Fourier-Stieltjes transform has also been discussed in \cite{KPG},  where the asymptotics was characterized by a similar, although not manifestly equivalent,  parameter.  The context of that work was  different,  involving relating fractal spectral measures in quantum mechanics to the large time asymptotics  of  of temporal correlation functions.
Translating to the present context,  it was shown that 
\begin{align}
\frac{1}{J}\int_0^J dJ'\, |Z(J')|^2 \sim J^{-{\cal D}_2}
\end{align}
in the limit of large $J$, where ${\cal D}_2$ is the correlation dimension of the fractal measure.
For the parameter $s_{\rm max}$ which we have chosen to characterize the asymptotic behavior,  the correspondence with another definition of dimension is unknown although, as noted above, $s_{\rm max}$ is bounded by the Haussdorf dimension.

For a simple choice of measures on the Cantor set, and generalized Cantor set, we will show that the bound $s_{\rm max}\le d$ is exactly saturated, so that $s_{\rm max}=d$. The convergence of $I_s$ for $s\le d$ is realized in non-trivial way, since $Z(\vec J)$ does not fall off with large $J$. Indeed, there is an infinite and unbounded sequence of values of $J$ having the same positive $Z(\vec J)$.  Numerical results for the characteristic function of the Lorenz attractor yields an estimate for $s_{\rm max}$, which is consistent with the known Hausdorff dimension.  It must be emphasized that this estimate is far too crude to claim equivalence with the Hausdorff dimension, as opposed to the slightly lower correlation dimension.

\section{Hopf characteristic functions for integer dimension}

At present, chaotic dynamical systems for which the characteristic function is known exactly have integer dimension $d$, and the invariant measure can be written as products of finite functions with delta functions. In this case it is not difficult to see that the maximum value of $s$ such that  \eqref{defcon} converges is $s_{\rm max}=d$.

An example of an exactly calculable characteristic function is that associated with the Orszag--Mclaughlin dynamical system \cite{Orszag},  defined by the equations of motion,
\begin{align}
\frac{dx_i}{dt}=x_{i+1}x_{i+2} + x_{i-1}x_{i-2}-2x_{i+1}x_{i-1}
\end{align}
with $i=1,\cdots,n$ and periodic identification $x_{i+n}=x_i$.  The Hopf characteristic function, computed in \cite{MarstonMa}, is given by,
\begin{align}
Z(\vec J)=\Gamma(n/2)(R|\vec J|/2)^{1-n/2}J_{n/2-1}(R|\vec J|)\, ,
\end{align}
where $R$ is a constant related to a conserved quantity of the motion, and $J_{n/2-1}$ is a Bessel function.
This characteristic function corresponds to a probability distribution
\begin{align}
P(\vec x)=\frac{1}{S_{n-1}}\delta(|\vec x|-R),
\end{align}
so that the dimension of the chaotic orbit is $d=n-1$.  For large $|\vec J|$,
\begin{align}
Z(\vec J)\approx \Gamma(n/2)\left(\frac{R|\vec J|}{2}\right)^{1-n/2}\sqrt{\frac{2}{R|\vec J|\pi}}\cos\left(R|\vec J|-\frac{(n/2-1)\pi}{2} - \frac{\pi}{4}\right).
\end{align}
The large $|\vec J|$ scaling, $|Z|^2 \sim |\vec J|^{1-n}$, implies convergence of \eqref{defcon} for $s_{\rm max}=n-1$, in agreement with the dimension $d=n-1$.

\section{Hopf characteristic function of the middle third Cantor set}

The middle third Cantor set is a fractal defined by starting with the unit interval and removing the middle third, and then the middle third of each remaining segment, ad infinitum.
The box counting dimension $D$ is defined by taking segments of length epsilon, and determining how the number of such segments necessary to cover the Cantor set scales as $\epsilon\rightarrow 0$. Taking $\epsilon=(1/3)^n$, it is not hard to see that the number N required to cover the set is $2^n$.  N scales like $(1/\epsilon)^D$ with $$D=\frac{\ln(2)}{\ln(3)}\, .$$  In this case the box counting dimension is equal to other common definitions of dimension, such as the Hausdorff dimension.

To compute the characteristic function $\left<\exp(iJX)\right>$, we define a measure on the Cantor set by  the $n\rightarrow\infty$ limit of a uniform probability distribution, with constant probability per unit length, on the set obtained after middle thirds have been removed $n$ times.
Writing the measure at each step as $d\mu(x)=\rho_n(x)dx$, one has
\begin{align}\rho_n(x)= \left(\frac{3}{2}\right)^n,\end{align}
such that $\int_{{\cal S}_n} dx \rho_n(x)=1$. Here ${\cal S}_n$ is the set obtained by removing the middle thirds $n$ times. The characteristic function is
\begin{align}\label{limit}
\lim_{n\rightarrow\infty} Z_n(J)\,
\end{align}
where
\begin{align}
Z_n(J)=\int_{{\cal S}_n} dx \rho_n(x)\exp(iJx)
\end{align}
While the $n\rightarrow\infty$ limit of $\rho_n$ does not exist, the $n\rightarrow\infty$ limit of $Z_n$ converges. Indeed, rapid convergence to a continuous function of is observed numerically.

\begin{centering}
\begin{figure}
\includegraphics[width=120mm]{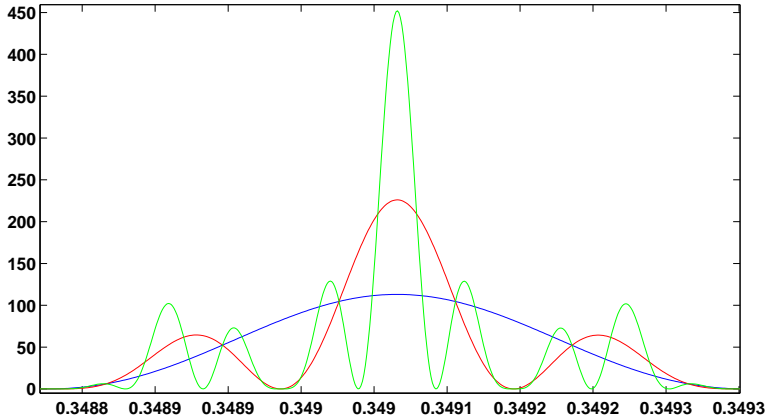}
\caption{\label{DistLim} Plots of $\chi_{8}(j)$ (blue), $\chi_{9}(j)$ (red) and $\chi_{10}(j)$ (green) in the neighborhood of $j=\pi/9$}
\end{figure}
\end{centering}

From the construction of the Cantor set, it is not hard to demonstrate that
\begin{align}\label{scaling}
Z_{n+1}(3J)=\frac{1}{2}\left(1+e^{2iJ}\right)Z_n(J)\, .
\end{align}
Existence of the limit \eqref{limit} and the scaling relation \eqref{scaling} implies
\begin{align}
Z(3J)=\frac{1}{2}\left(1+e^{2iJ}\right)Z(J)\,
\end{align}
Note $Z(3^m \pi n)$ is independent of $m$, where $m$ and $n$ are is integers.  One can check numerically that $J=\pi n$ is not necessarily a zero of $Z$.  Thus, it is not true that $\lim_{J\rightarrow\infty}Z(J)=0$, which would hold if the invariant measure could be written in terms of a finite probability density function.  Perhaps contrary to expectation, $Z(J)$ does not obey any simple large $J$ asymptotics, such as a power law fall off.
The scaling relation \eqref{scaling} implies
\begin{align}
\int_{3\tilde J}^{9\tilde J}\,dj\,|Z(j)|^2j^{-\alpha} &= 3^{1-\alpha}\int_{\tilde J}^{3\tilde J}\,dj\,|Z(3j)|^2 j^{-\alpha} \nonumber\\ &=
\frac{1}{2}3^{1-\alpha}\int_{\tilde J}^{3\tilde J}dj\,(1+\cos(2j))|Z(j)|^2j^{-\alpha}\, ,
\end{align}
for any $\tilde J$.
Thus,
\begin{align}\label{sumrep}
\int_{\tilde 3J}^{\infty}|Z(j)|^2 j^{-\alpha} = \sum_{n=1}^\infty \int_{\tilde J}^{3\tilde J}\, dj\, \left(\frac{3^{1-\alpha}}{2}\right)^n \chi_n(j) \, |Z(j)|^2  j^{-\alpha}\, ,
\end{align}
where
\begin{align}\label{chidef1} \chi_n \equiv \prod_{m=1}^n \left(1+\cos(3^{m-1}2j)\right)\, .
\end{align}
 Note that $\chi_n(j)$ does not converge to any function in the large $n$ limit. In particular, it diverges for any $j=\frac{\pi l}{3^k}$ with integer $l$ and positive integer $k$. These values of $j$ are a dense subset of the reals. 
Figure \ref{DistLim} shows $\chi_8,\chi_9$ and $\chi_{10}$ in the neighborhood of $j=\pi/9$.  Despite the lack of convergence to any function, one can show that the $n\rightarrow\infty$ limit is a distribution, such that the limit
\begin{align}
\lim_{n\rightarrow\infty}\int_{\tilde J}^{3\tilde J}\, dj\,\chi_n(j)f(j) = (\chi,f)
\end{align}
exists for any function $f(j)$ with a uniformly convergent Fourier series expansion\footnote{We do not have a proof of convergence when the Fourier series is not uniformly convergent.} on the interval $[\tilde J,3\tilde J]$. The proof is given in appendix A. Rapid numerical convergence of $(\chi_n,f)$ for some arbitrarily chosen test functions is shown in table 1.  Assuming that $\lim_{n\rightarrow\infty}\chi_n(j)$ converges to a distribution sufficiently fast, then the sum \eqref{sumrep} behaves as
\begin{align}
\sum_{n} \left(\frac{3^{1-\alpha}}{2}\right)^n\,(\chi, |Z(j)|^2  j^{-\alpha})
\end{align} for large $n$.
Convergence of this sum requires $\frac{3^{1-\alpha}}{2}<1$.  Therefore the maximum value of $s$ such that \eqref{defcon} converges\footnote{One could also have attempted to apply the scaling relation \eqref{scaling} to determine convergence of $\int_0^{3^K \tilde J}\, dJ\, |Z(J)|^2 J^{-\alpha}$ for integer $K$ as $K\rightarrow\infty$.  However this approach requires considerable care, as it involves evaluating $(\chi_K,f)$ with an integration over a region bounded by $J=0$, where $f=|Z(J|^2J^{-\alpha}$ has a singularity.}, is \begin{align}s_{\rm max}=1-\alpha=\frac{\ln(2)}{\ln(3)},\end{align} in agreement with the  Hausdorff dimension.

\begin{table}
\begin{center}
\begin{tabular}{c|c|c}n & $(\chi_n,f_1)$ & $(\chi_n, f_2)$ \\ \hline
1 & 0.36677117 & -0.05595122\\ \hline
2 & 0.43466270 & -0.12962766\\ \hline
3 & 0.43529857 & -0.02549191\\ \hline
4 & 0.43545871 & -0.03928547\\ \hline
5 & 0.43546377 & -0.03951139\\ \hline
6 & 0.43546202 & -0.03948616\\ \hline
7 & 0.43546384 & -0.03948616\\ \hline
8 & 0.43546442 & -0.03986482\\ \hline
9 & 0.43546424 & -0.03987476 \\ \hline
10 & 0.43546423& -0.03987374\\ \hline
11 & 0.43546423& -0.03987312\\ \hline
12 & 0.43546423& -0.03987311\\ \hline
13 & 0.43546423& -0.03987305\\ \hline
14 & 0.43546423& -0.03987306\\ \hline
\end{tabular}
\caption{The integral $\int_{1}^{3} dJ\, \chi_n(J) f(J)$ evaluated for arbitrarily chosen test functions $f_1(J)=\exp\left( -5(J-2)^2 \right) $ and $f_2=\cos(20 J)$. Note the rapid convergence with increasing $n$.}
\end{center}
\end{table}


\section{Hopf characteristic function of the generalized Cantor set}

A correspondence between the asymptotic behavior of the characteristic function and the fractional dimension also holds
for the generalized Cantor set.
The generalized Cantor set can be defined as follows.
Consider the map \begin{align}\label{map}
x\rightarrow &2\eta x\,\,{\rm for}\,\, x<\frac{1}{2} \nonumber \\
x\rightarrow &1-2\eta(1-x)\,\,{\rm for}\,\, x>\frac{1}{2}
\end{align} for $\eta>1$.
With each application of the map,  points between $\frac{1}{2\eta}$ and $1-\frac{1}{2\eta}$ are mapped outside the unit interval $[0,1]$
while the segments $[0,\frac{1}{2\eta}]$ and $[1-\frac{1}{2\eta},1]$ are stretched to fill the interval $[0.1]$.
The generalized Cantor set is the set of initial conditions on the interval $x=[0,1]$ which are not mapped outside the interval after any number of applications of \eqref{map}.  For $\eta=\frac{3}{2}$, this set is the middle third Cantor set discussed above.

The characteristic function of the generalized Cantor set can be studied in essentially the same manner as the middle third Cantor set.  Starting with the distribution $\rho=1$ on the unit interval,  a new distribution can be defined with each application of the map \eqref{map}.  With each application of the map, the set of initial conditions which remain within the unit interval is reduced, and the distribution is taken to be uniform (constant) on the remaining set of inital conditions,  integrating to 1.  For instance, after the first application of the map \eqref{map}, the surviving initial conditions are the segments $[0,\frac{1}{2\eta}]$ and $[1-\frac{1}{2\eta},1]$, on which the distribution is $\rho=\eta$.
After $n$ applications of the map, one has \begin{align}
\rho_n=\eta^n
\end{align}
on $2^n$ segments ${\cal S}_n$  within the unit interval.  The characteristic function is
\begin{align}
Z_n(J)=\int_{ {\cal S}_n} dx \rho_n \exp(iJx)
\end{align}

\begin{table}
\begin{center}
\begin{tabular}{c|c|c}n & $(\chi_{\eta,n},f_1)$ & $(\chi_{\eta,n}, f_2)$ \\ \hline
1 & 0.07620179 & -0.04757256 \\ \hline
2 & 0.11068899 & -0.01065505\\ \hline
3 & 0.08751986 & -0.00348152\\ \hline
4 & 0.08803434 & -0.00178861\\ \hline
5 & 0.08802082 & -0.00481686\\ \hline
6 & 0.08802103 & -0.00479188\\ \hline
7 & 0.08802102 & -0.00479230\\ \hline
8 & 0.08802103 & -0.00479224\\ \hline
9 & 0.08802103 & -0.00479226\\ \hline
10 & 0.08802103& -0.00479226 \\ \hline
11 & 0.08802103& -0.00479226 \\ \hline
12 & 0.08802103& -0.00479226\\ \hline
13 & 0.08802103& -0.00479226\\ \hline
14 & 0.08802103& -0.00479226 \\ \hline
\end{tabular}
\caption{The integral $\int_{1}^{2\eta} dJ\, \chi_{\eta,n}(J) f(J)$ evaluated for $\eta=5/4$ and test functions $f_1(J)= \exp\left( -5(J-2)^2 \right)$ and $f_2=\cos(20 J)$. Note the rapid convergence with increasing $n$.}
\end{center}
\end{table}

The definition of ${\cal S}_n$ and $\rho_n$ yields the relation,
\begin{align}
Z_{n+1}(2\eta J)=\frac{1}{2}\left(1+e^{i2\eta J(1-\frac{1}{2\eta})}\right)Z_n(J)\, .
\end{align}
Therefore, existence of the limit $Z(J)=\lim_{n\rightarrow\infty}Z_n(J)$ implies;
\begin{align}
Z(2\eta J)=\frac{1}{2}\left(1+e^{i2\eta J(1-\frac{1}{2\eta})}\right)Z(J)
\end{align}
Thus,
\begin{align}
\int_{2\eta\tilde J}^{(2\eta)^2\tilde J}\,&dj\,|Z(j)|^2j^{-\alpha} = (2\eta)^{1-\alpha}\int_{\tilde J}^{(2\eta)\tilde J}\,dj\,|Z(2\eta j)|^2 j^{-\alpha} \nonumber\\ =
&\frac{1}{2}(2\eta)^{1-\alpha}\int_{\tilde J}^{(2\eta)\tilde J}dj\,\left(1+\cos\left(2\eta J(1-\frac{1}{2\eta})\right)\right)|Z(j)|^2j^{-\alpha}\, ,
\end{align}
for any $\tilde J$.
Therefore
\begin{align}\label{sumrep2}
\int_{2\eta\tilde J}^{\infty}\, dj\,|Z(j)|^2 j^{-\alpha} = \sum_{n=1}^\infty \int_{\tilde J}^{2\eta\tilde J}\, dj\, \left(\frac{(2\eta)^{1-\alpha}}{2}\right)^n \chi_n(j) \, |Z(j)|^2  j^{-\alpha}\, ,
\end{align}
where
\begin{align}\label{chidef2} \chi_{\eta,n} \equiv \prod_{m=1}^n  \left(1+\cos\left((2\eta)^{m-1}2\eta J(1-\frac{1}{2\eta})\right)\right)       \, .
\end{align}
The $n\rightarrow\infty$ limit of $\chi_{\eta,n}$ does not converge to a continuous function.  However, we conjecture that it converges sufficiently rapidly to a distribution, such that the sum \eqref{sumrep2} converges when $\frac{(2\eta)^{1-\alpha}}{2}<1$.  For several values of $\eta$ and for several arbitrarily chosen test functions $f$, we have numerically checked that $(\chi_{\eta,n},f)$ converges rapidly with increasing $n$, as illustrated in table 2. The maximum value of $s$ such that \eqref{defcon} converges, is then
\begin{align}
s_{\rm max}=1-\alpha=\frac{\ln(2)}{\ln(2\eta)},
\end{align}
which is equal to the Hausdorff dimension.


\section{Hopf characteristic function of the Lorenz attractor}

The Cantor sets considered above are self similar fractal sets, for which scaling arguments yield information about the asymptotic behavior of the characteristic function. The chaotic invariant sets associated with continuous dynamical systems are not generally self similar, making it difficult to determine the large $J$ behavior of $Z(\vec J)$ analytically. In the following we give crude estimates for the asymptotic behavior of the Lorenz attractor, i.e. for the maximum value of $s$ such that \eqref{defcon} converges, by a numerical computation of $Z(\vec J)$.  This approach is hampered by the fact that one can only compute $Z$ up to a finite value of $J$, limited in size by the available computational power.

The Lorenz attractor \cite{Lorenz} is defined by
\begin{align}
\frac{dx}{dt}=\sigma(y-x), \qquad
\frac{dy}{dt}=x(\rho-z)-y, \qquad
\frac{dz}{dt}=xy-\beta z.
\end{align}
We choose the value $\sigma=10,\,\rho=28,\, \beta=8/3$, and a set of $N$ distinct initial conditions within the basin of attraction. These are evolved for a fixed time $T$. The characteristic function is given by
\begin{align}Z(\vec J)=\lim_{T\rightarrow\infty}
=\frac{1}{N}\sum_{i=1}^N \frac{1}{T}\int_0^T dt\, e^{i\vec J\cdot\vec x_i(t)}
\end{align}
The choice of $N$ is irrelevant; it may just as well be set to one. However it is computationally most efficient to choose a large $N$ and evolve each set of initial conditions in parallel for a finite time to estimate $Z(\vec J)$.
As the value of $|\vec J|$ is increased, one must increase $N$ and/or the
density of sampling of $\vec x(t)$ in time to avoid a problem akin to aliasing\footnote{The rate at which sampling densities must be adjusted with increasing $|\vec J|$ is presumably related to the box counting dimension.}.

\begin{centering}
\begin{figure}[t]
 \includegraphics[width=\textwidth]{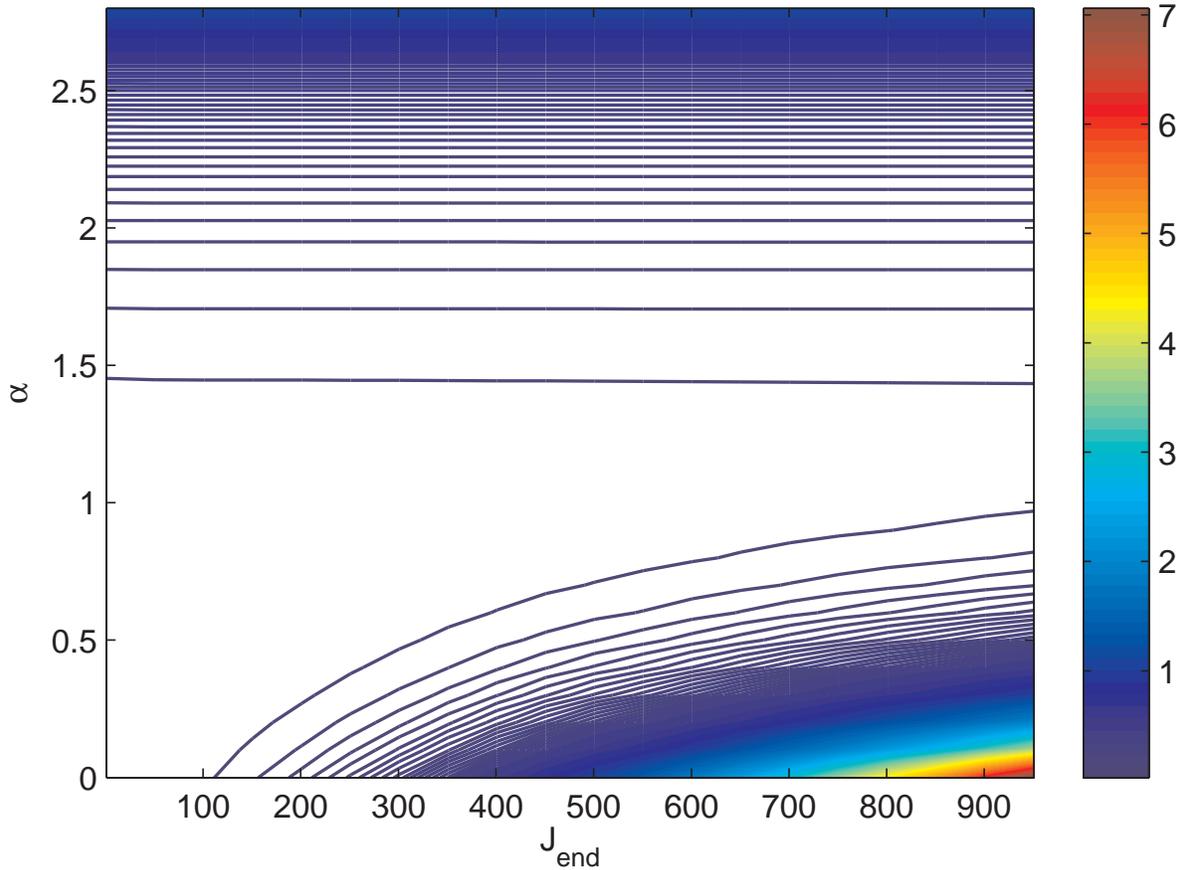}
 \caption{\label{LorenzConv} Contour plot of the integral $\int_\epsilon^{J_{\rm end}} dj\, j^{2-\alpha}|Z(j\hat z)|^2$ as a function of $\alpha$ and $J_{\rm end}$.}
\end{figure}
\end{centering}

We wish to determine the minimum value of $\alpha=3-s$ such that the integral,
\begin{align}
I=\int d^3J\, |\vec J|^{-\alpha}|Z(\vec J)|^2
\end{align}
converges
Because of computational constraints, we will instead pick a three-dimensional unit vector $\hat n$ and demand convergence of the integral
\begin{align}\label{ConvTest}
I(\hat n)=\int_{0}^{\infty} dJ\,J^{2} J^{-\alpha}|Z(J\hat n)|^2
\end{align}
For the few choices of $\hat n$ we have made, the result seems to be insensitive to the choice of $\hat n$.
A contour plot of
the integral \eqref{ConvTest} as a function of $\alpha$ and the endpoint of integration $J_{\rm end}$ is shown in figure \ref{LorenzConv}, suggesting  a critical $\alpha$ in the neighborhood of $\alpha=1$, above which the integral converges and below which it diverges.  This is not inconsistent with saturation of the bound on the dimension, $d>3-\alpha_{\rm crit}$, since the Hausdorff dimension (see \cite{Viswanath} for a recent estimate) is  $d\approx 2.06$.

Of course,  numerical results are extremely crude and merely suggestive of asymptotic properties, since the they are limited to finite $J$ endpoints of integration.  A more clever approach is required to accurately determine the asymptotics of $Z(\vec J)$.  It would be very interesting if there existed some asymptotic analysis of the differential equations which the characteristic function satisfies (see e.g. \cite{MarstonMa}), which could determine $\alpha_{\rm crit}$.  As in the case of the Cantor set, there is no reason to expect the large $J$ behavior of $Z(\vec J)$ to a have a simple functional form, such as a power law fall off.  Thus it may be that one must consider some suitably averaged, or integrated form of these differential equations to obtain $\alpha_{\rm crit}$.

\section{Conclusions}

While the small $J$ behavior of the Hopf characteristic function $Z(\vec J)$ is related to correlation functions, via the coefficients of the Taylor--Mclaurin expansion, the large $J$ behavior is related to dimensionality.
For the parameter we have chosen to measure asymptotics, the large $J$ behavior of the characteristic function satisfies a bound given by the Hausdorff dimension of the set.  We showed this bound to be saturated for a natural measure on the generalized Cantor set.  Crude numerical results are also consistent with saturation for the Lorenz attractor.  
A number of interesting questions remain un-answered.

In order to make the arguments relating the dimensionality of the Cantor set to the asymptotic behavior of $Z(J)$ more rigorous, it remains only to prove that that the functions $\chi_n$, defined in \eqref{chidef1} and \eqref{chidef2} approach a limiting distribution sufficiently fast at large $n$. Thus far we have only shown (see appendix A) that $\lim_{n\rightarrow\infty}\chi_n$ is a distribution on the space of functions which have absolutely convergent Fourier series expansions. A better understanding of the properties of the functions $\chi_n$ would be valuable.  Note that they do bear some resemblance to Weierstrass's non-differentiable function.

Subject to certain assumptions about analyticity, the small $J$ behavior of $Z(\vec J)$ is related to the large $J$ behavior.  In principle, the large $J$ behavior could be determined by a re-summation of the cumulant expansion $\ln(Z)=\sum_{n=0}^\infty c_nJ^n$.  The dimensionality should give constraints on the behavior of the cumulants $c_n$ for large $n$.   It could be very interesting to elucidate these constraints.

Perhaps the differential equations satisfied by the Hopf characteristic function of a chaotic dynamical system, which follow from the equations of motion, can somehow be used to draw conclusions about the large $J$ asymptotics. If so, such an approach will likely involve an integrated or suitably averaged form of these equations, since $Z(\vec J)$ need not display a simple asymptotic fall off when the dimension is non-integer.  Indeed, we have seen that there is no asymptotic fall off for the Cantor set, nor do we observe a power law fall off for the Lorenz attractor.

This work was motivated in part by \cite{G,CGG}, in which a class of chaotic dynamical systems was found for which exact statistics, in the form of a probability distribution function, is known.  These dynamical systems were reverse engineered starting with a probability  distribution and a two-form. Due to the existence of a finite probability distribution function, these systems do not have fractional dimension, and are dissipative in some regions of the chaotic orbit but not globally.  There may be a way to extend the inverse approach to dissipative systems with fractional dimensions, starting with a characteristic function having no Fourier transform, rather than starting from a probability distribution.

\section{Acknowledgements}
Cengiz Pehlevan would like to thank the Swartz foundation for support
through a fellowship. G.S Guralnik is supported in part by the
U.S. Department of Energy (DOE) under DE-FG02-91ER40688-Task D. We thank the referee for pointing out  reference \cite{KPG} to us.

\appendix

\appsection{}

Here we show that the $n\rightarrow\infty$ limit of
\begin{align} \chi_n \equiv \prod_{m=1}^n \left(1+\cos(3^{m-1}2j)\right)\, .
\end{align}
is a distribution, such that the limit
\begin{align}\label{distrib2}
\lim_{n\rightarrow\infty}\int_{\tilde J}^{3\tilde J}\, dj\,\chi_n(j)f(j)
\end{align}
exists for any function $f(j)$ with a uniformly convergent Fourier series expansion on the interval $[\tilde J,3\tilde J]$.

In light of the identity
\begin{align}\label{trig}
\cos(a)\cos(b)=\frac{\cos(a+b)+\cos(a-b)}{2},
\end{align}
the highest order term in the Fourier expansion of $\chi_n$ is $\cos(pj)$ with
\begin{align}
p=2\sum_{m=1}^{n}3^{m-1}=3^{n-1}
\end{align}
It follows from \eqref{trig} that the Fourier expansion of $\chi_{n+1}=(1+\cos(3^n2j))\chi_n$ will have the same coefficients as that of $\chi_{n}$ for all modes of order $p$ or less, with higher order modes having coefficients less than or equal to the maximum coefficient of $\chi_n$, which by induction is 1. Therefore $\chi_n$ has a Fourier expansion in cosines of even integer multiples of $j$ with positive coefficients less than or equal to one.

For simplicity, let us now choose $\tilde J=\pi/2$ and exchange $f(j)$ in \eqref{distrib2} with a $\pi$ periodic function which is equivalent to $f(j)$ in the integration range.  Now $f^P(j)$ and $\chi_n$ have Fourier expansions
\begin{align}
f^P(j)=\sum_{k=0}^{\infty}\left(a_k \cos(2kj)+ b_k \sin(2kj)\right),\nonumber\\ \chi_n(j)=\sum_{k=0}^{3^n-1} c_k \cos(2kj),\qquad 0\le c_k \le 1,
\end{align}
such that
\begin{align}\label{convt}
\int^{3\pi/2}_{\pi/2} dj\, f^P(j)\chi_n(j)=\pi a_0 c_0 + \frac{\pi}{2}\sum_{k=1}^{3^n-1}a_k c_k.
\end{align}
If the Fourier series expansion of $f_P(j)$ is absolutely convergent, i.e.
\begin{align}
\sum_{k=0}^{\infty}|a_k|<\infty
\end{align}
converges, then by the comparison test
\begin{align}
\lim_{n\rightarrow\infty}\sum_{k=0}^{3^n-1}|a_k c_k|
\end{align}
must also converge. Thus \eqref{convt} converges in the limit $n\rightarrow\infty$, and we conclude that $\lim_{n\rightarrow\infty}\chi_n$ is a distribution.

\end{document}